\documentclass[pdflatex,sn-mathphys]{sn-jnl}

\jyear{2022}%

\theoremstyle{thmstyleone}%

\theoremstyle{thmstyletwo}%

\theoremstyle{thmstylethree}%

\begin{document}

\title[AAR in East Antarctica: Temperature and thinning]{Apparent ice accumulation rate in East Antarctica: Relation with temperature and thinning pattern}

\author*[1]{\fnm{Radhendushka} \sur{Srivastava}}\email{rsrivastava@iitb.ac.in}

\author[2]{\fnm{Debasis} \sur{Sengupta}}\email{sdebasis@isical.ac.in}

\author[3]{\fnm{Prosenjit} \sur{Ghosh}}\email{pghosh@iisc.ac.in }

\affil*[1]{\orgdiv{Department of Mathematics}, \orgname{Indian Institute of Technology Bombay}, \orgaddress{ \city{Mumbai}, \postcode{400076}, \state{Maharashtra}, \country{India}}}

\affil[2]{\orgdiv{Applied Statistics Unit}, \orgname{Indian Statistical Institute}, \orgaddress{ \city{Kolkata}, \postcode{700108}, \state{West Bengal}, \country{India}}}

\affil[3]{\orgdiv{Centre for Earth Science}, \orgname{Indian Institute of Science}, \orgaddress{\city{Bangalore}, \postcode{560012}, \state{Karnataka}, \country{India}}}

\abstract{
We present here formal evidence of a strong linkage between temperature and East Antarctic ice accumulation over the past eight hundred kiloyears, after accounting for thinning. The conclusions are based on statistical analysis of a proposed empirical model based on ice core data from multiple locations with ground topography ranging from local peaks to local valleys. The method permits adjustment of the apparent accumulation rate for a very general thinning process of ice sheet over the ages, is robust to any misspecification of the age scale, and does not require delineation of the accumulation rate from thinning. 
Records show 5\%\ to 8\%\ increase in the accumulation rate for every 1${}^\circ$C rise in 
temperature. This is consistent with the theoretical expectation on the average rate of increase in moisture absorption capacity of the atmosphere with rise in temperature, as inferred from the Clausius-Clapeyron equation. This finding reinforces indications of the resilience of the Antarctic Ice Sheet to the effects of warming induced by climate change, which have been documented in other studies based on recent data. Analysis of the thinning pattern of ice revealed an exponential rate of thinning over several glacial cycles and eventual attainment of a saturation level.}


\maketitle

\section{Introduction}\label{sec1}
In paleo-climatic studies, analysis of data from drilled ice cores forms a rich source of information about the past climate patterns. Direct or indirect records of ambient temperature and chemical composition of air are preserved in the layers of these ice cores. Several paleo-climatic variables like temperature, carbon-di-oxide etc. are obtained by using stable isotopes and analysis of trapped gaseous molecules preserved in layers of ice, respectively. Researchers have used the information obtained from ice cores to study the long-term climate patterns like glacial-interglacial cycles, which allow verification of the Milankovitch frequencies depicting the changes in the sun-earth configuration, affecting global temperature and levels of greenhouse gases in the global atmosphere (see, e.g., \cite{Rapp2019,Jouzel2007,Petit1999,Siegenthaler1987}). The net effect of the present phase of warming is unprecedented; with significant loss of ice sheet as the amount of ice gained by snowfall is found smaller compared with the loss of ice due to iceberg melting. Ideally, such systems attain equilibrium when gain and loss are nearly equal, suggesting no major change in the ice mass. Independent studies have also shown that the rise in the amount of snowfall could stall the projected sea level rise, but it is far from enough to bring the system back into equilibrium.

The physical mechanism that can explain increased snowfall is that the presence of excess water vapor in the atmosphere, as the planet gets warmer, is expected to bring more precipitation in the form of snowfall. The moisture holding capacity of global atmosphere increases by about 7\%\ with 1$^\circ$C (1.8$^\circ$F) rise in global temperature as evident from the Clausius-Clapeyron equation \cite{Algarra2020}. The Gomez ice core record from western Antarctica reveals a doubling of accumulation since the 1850s \cite{Thomas2008}, consistent with the upward trend reported in the regional precipitation for the period 1980–2004 \cite{vandenBroeke2006} and satellite altimeter based assessment indicating an increase in elevation in the western Peninsula for 1992–2003, as an aftermath of greater volume of snowfall \cite{Wingham2006}.
These evidences of excess snow/ice accumulation overtaking the melting of ice challenge the conventional wisdom of mass imbalance.  Such a mechanism of excess snow accumulation in a warmer world, if found to hold over a longer period of time, would demonstrate the response of the hydrological system to maintain the thermostat of the Earth in scenarios of global warming. We explore here apparent ice accumulation rate during paleo time scale from Antarctic ice core records, where independent knowledge of the exact timing of ice deposition (from matching the solar frequencies) provide a unique opportunity to correlate temperature with thickness of ice layers at the corresponding ages.

The relationship between temperature and apparent ice accumulation rate has been a subject matter of scientific interest for many years. A study of the Guliya ice-core record of the past 300 years on the crest of the Kunlun Shan, Central Asia \cite{Yang2006} revealed a relationship between ice accumulation rate and temperature. The temperature and precipitation rate in the region are strongly correlated, with the observed excess precipitation rate lagging behind that of temperature maxima by 20 to 40 years. A study of the relationship between ice accumulation rate and temperature for the past 31,000 years, on the basis of ice core data from the WDC site of West Antarctica, also showed a consistent relationship between temperature and apparent ice accumulation rate \cite{Fudge2016}. However, an inverse relationship between apparent ice accumulation rate and temperature during the past seven millenia was found from the analysis of the Green Ice Sheet Project 2 (GISP2) data in central Greenland \cite{Cuffey1997}. This finding is indicative of probable thinning of the ice sheet during warmer climates. While the evidence from West Antarctica and elsewhere during that period goes in the opposite direction, there is no comparable study from East Antarctica. Marine sediment records suggest that the East Antarctic ice sheet may have experienced occasional retreat from coastal areas during the Pliocene epoch, 5.3 million to 2.6 million years ago \cite{Cook2013}. It may be noted that during the warmer phases of this interval, the global average temperatures were 2 to 3 degrees Celsius warmer than in pre-industrial times \cite{Robinson2008}, and carbon dioxide level were similar to the modern day level \cite{delaVega2020}. The extent of such retreat has been questioned in subsequent research \cite{Shakun2018}. 

\begin{figure}[!h]
\centering
\includegraphics[width=2.5in,height=2in]{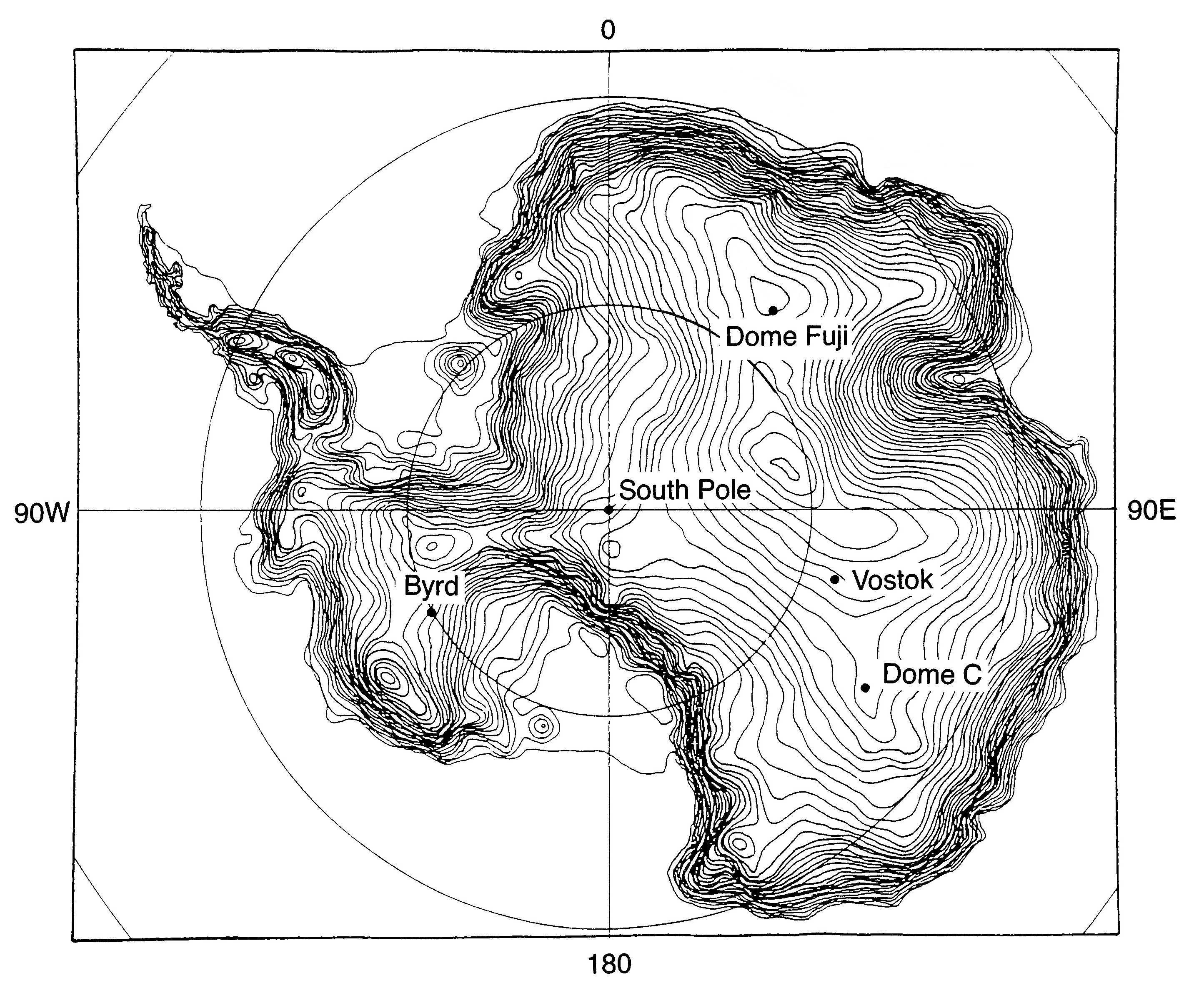}\caption{ The locations of Dome Fuji, Lake Vostok and Dome C are shown on the Antarctic continent (Figure derived from Watanabe et al., 2003).}
\label{Fig1}
\end{figure}

This makes the East Antractic ice sheet a good archive for probing the ice accumulation rate until the Pliocene epoch. Using ice core data from East Antarctica, publicly available at the website of National Center for Environmental Information (www.ncdc.noaa.gov), we present here a pattern of ice accumulation over an age span of several hundred thousand years, and establish a relationship with temperature. The three data sets, with overview given in Table~\ref{DATA_info}, consist of age (measured according to the AICC2012 scale~\cite{Bazin2013} in Kilo Years Before Present or KYrBP, the “present” being the year 1950) of slices of ice at various depths of the ice core, the annual average sea surface temperature anomaly (reconstructed from oxygen isotope used as proxy and expressed in degree Celsius as deviation from long term average) at that age, and other parameters that we do not use here. The sites of data collection are shown in Figure~\ref{Fig1}.                         
\begin{table}[h]
  \begin{center}
  \caption{Overview of three ice core data sets from East Antarctica}\label{DATA_info}
  \medskip
  \begin{tabular}{p{0.5in}p{0.5in}p{0.52in}p{0.47in}p{0.5in}p{0.5in}p{0.5in}}
  \hline
 Name &Location &Age range (KYrBP) &Depth range (meters) &Ice sheet altitude above average sea level (meters asl) &Sample size &Sampling interval (meters)\\
\hline
\phantom{phantom} Dome &$77^{\circ}$19'S $39^{\circ}$42'E&$-0.061$ to 298.111 &0 to\phantom{phan} 2403.5 &\phantom{phantom} 782 to &\phantom{phantom} 1223 &\phantom{phantom} $1.97^*$\\[.1cm]
Fuji &$77^{\circ}$30'S $37^{\circ}$30'E & 298.165 to 715.892 &2403.1 to 3028.1 & 3810 &6224 &0.10\\[.6cm]
Lake \ \ Vostok&$77^{\circ}$50'S $106^{\circ}$00'E	&$-0.005$ to\phantom{phan} 403.780	&0 to 3263 &$-500$ to\phantom{p} 3488 &3311 &1.00\\[.6cm]
EPICA Dome C &$75^{\circ}$06'S $123^{\circ}$20'E &$0.031$ to 801.588 &6.6 to\phantom{phan} 3189.45\phantom{phan} &$-40$ to\phantom{ph} 3233 &5800 &0.55\\
\hline
  \end{tabular}\\
  \end{center}
${}^*${average sampling interval; actual sampling intervals are non-uniform.}
\end{table}

The age difference at successive depths indicates the number of years of accumulation represented by a particular range of depths. Here, instead of yearly ice layer thickness or accumulation rate (which is generally estimated at higher granularity through isotopes \cite{Cauquoin2015} or relationship with other factors\cite{Bazin2013}) we prefer to introduce a new term called “Apparent accumulation rate” of ice. We define the apparent accumulation rate (AAR) as the ratio of differences in the depth of successive ice core samples (measured in meters) and the age difference (in KYrBP) of these samples. The AAR is different from the conventional term ``ice layer thickness". It describes the ice accumulation per kilo year as it appears on site today, resulting from the net effect of precipitation, physical change of state like sublimation, melting (which should be minimal for the location and time interval of the present study), evaporation, lateral transport and also compaction due to strain from the overlying weight of stratified ice. 

\begin{figure}[t]
\centering
\begin{tabular}{ccc}
{\underline{\small Dome Fuji}}& {\underline{\small Lake Vostok}}& {\underline{\small EPICA Dome C}}\\
\includegraphics[width=1.4in,height=1.9in]{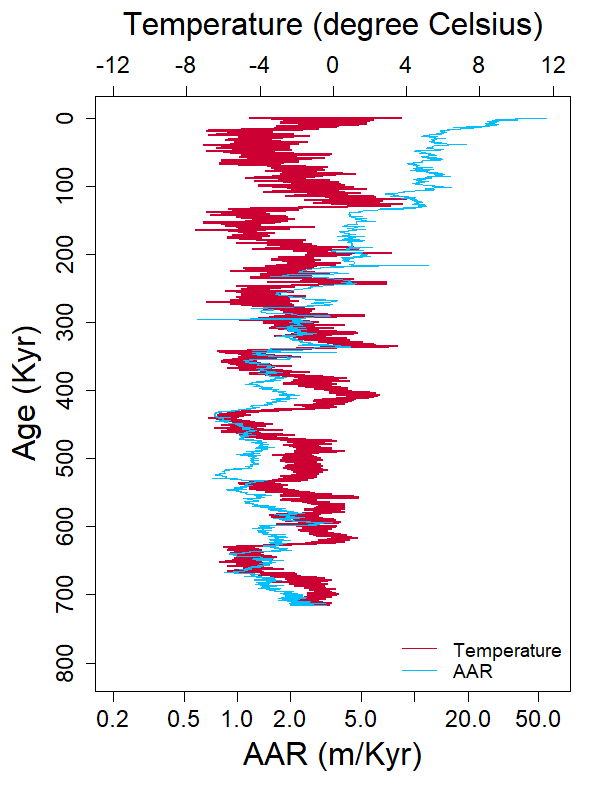}&
\includegraphics[width=1.4in,height=1.9in]{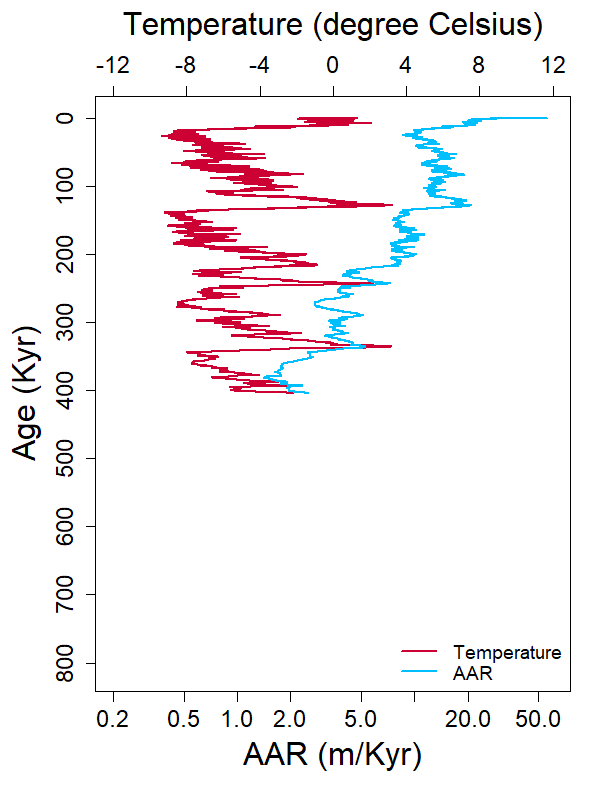}&
\includegraphics[width=1.4in,height=1.9in]{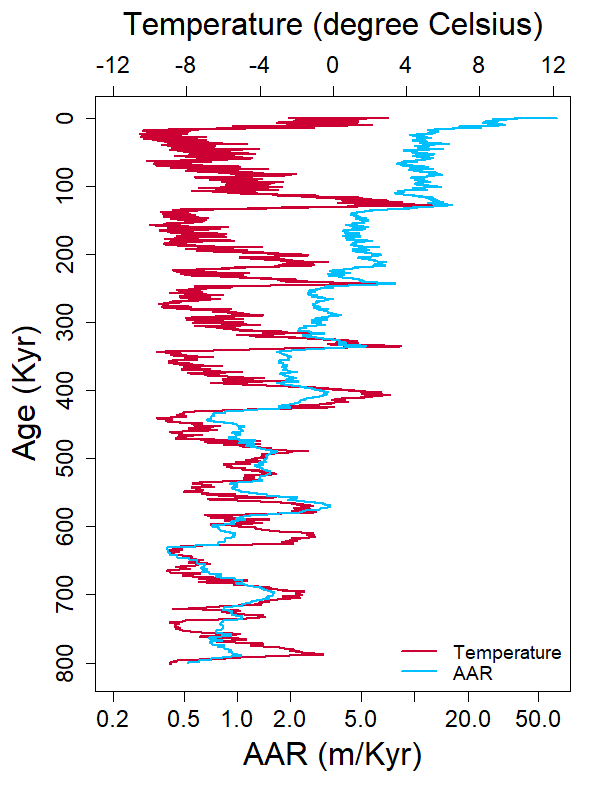}\\
\end{tabular}
\caption{The plots show the graphs of temperature anomaly (${}^\circ$C in red solid line) labeled along the top horizontal axis and apparent ice accumulation rate (AAR) (m/KYr in blue solid line) in log scale labeled along the bottom horizontal axis, versus age (KYrBP) labeled along left vertical axis.}
\label{Fig2}
\end{figure}

We study AAR as a function of age (in KYrBP) and compare it with the temperature anomaly (that is the estimate of temperature difference from average of last 1000 years in degrees Celsius) at those ages. Figure~\ref{Fig2} shows this relationship comparatively for the data of Dome Fuji, Lake Vostok and EPICA Dome C, from left to right. The scales of temperature anomaly, AAR (in log scale) and age are given on lower, upper and left axes, respectively. It can be seen that even though there is an overall decreasing trend in AAR over the wide range of age, its movements follow those in temperature. Wherever the temperature has a peak, the AAR generally has a peak, and vice versa. The movement of accumulation rate is in the direction of that of temperature, which is in agreement with what had been observed from the ice core data of the past 31 kilo years from the WDC site of West Antarctica \cite{Fudge2016}. It may be noted that a positive relationship between accumulation rate and temperature was {\it presumed} in \cite{Kahle2021} for simultaneously reconstructing temperature and accumulation rate.

Therefore, we proceed to describe a long-term relationship between AAR with estimated air temperature and relate our findings with existing knowledge of climatic shifts. We delineate the actual rate of accumulation of ice from the thinning process only for a descriptive purpose, and refrain from this distinction at the time of estimating the said relationship. This restraint, as well as a general framework that is resilient to misspecification of the time scale, allows us to make precise statistical inference on the effect of temperature.

\section{Results and Discussion}\label{Sec2}
\subsection{Effect of temperature on AAR}\label{SS2.1}
In view of the discussion of the previous section, we model AAR at temperature $x$ and age $z$ as a noisy version of the product of two components:
\begin{equation}
b(x,z)\approx (1+\gamma x)g(z),    
\label{eqn1}
\end{equation}
where, $\gamma$ is a constant parameter and $g$ is a decreasing function. The details of the model, including precise definitions, interpretation and antecedents are given in Section~\ref{Sec3}. We fit this model to the three data sets separately for three different age segments: 0 to 125 KYrBP (younger age of ice), 125 to 410 KYrBP (age of ice in the middle range) and 410 KYrBP to 800 KYrBP (older age of ice). Note that the age 125 KYrBP is approximately the crest of warming in between the two most recent glacial periods, and 410 KYrBP is approximately the crest of warming in between the fourth and the fifth most recent glacial periods. Data for the older age is not available for Lake Vostok. 

\begin{figure}[h]
\centering
\begin{tabular}{ccc}
{\underline{\small Dome Fuji}}& {\underline{\small Lake Vostok}}& {\underline{\small EPICA Dome C}}\\
\includegraphics[width=1.4in,height=1.9in]{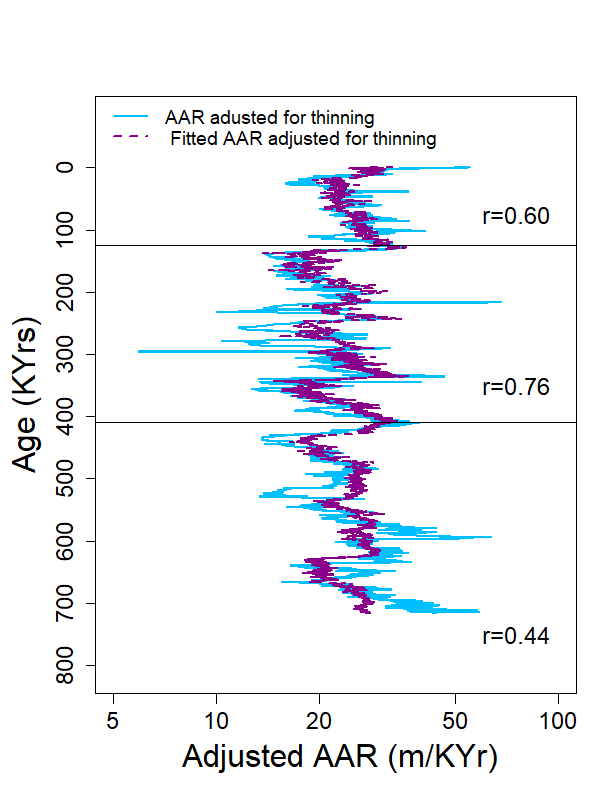}&
\includegraphics[width=1.4in,height=1.9in]{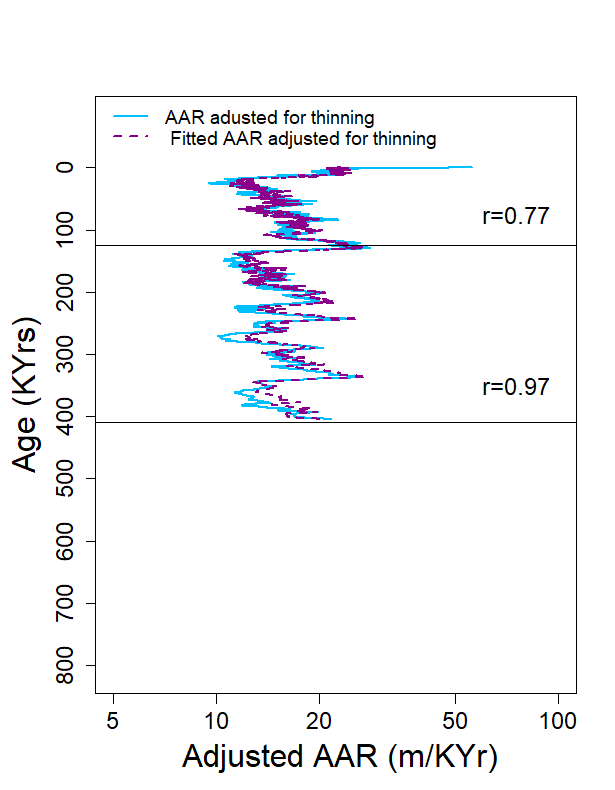}&
\includegraphics[width=1.4in,height=1.9in]{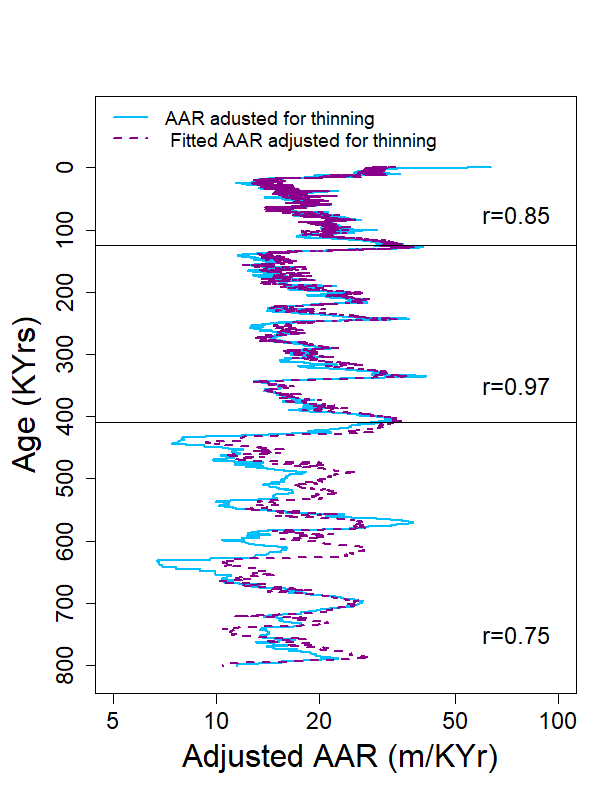}\\
\end{tabular}
\caption{The plots show the graphs of observed AAR adjusted for thinning (Solid blue line) and fitted AAR adjusted for thinning (dashed purple line), in log scale labeled along the bottom horizontal axis, versus age labeled along the left vertical axis. The two horizontal black lines correspond to ages 125 KYrBP and 410 KYrBP.}
\label{Fig3}
\end{figure}

Figure~\ref{Fig3} shows the plot of observed AAR adjusted for thinning (obtained by multiplying $b(x,z)$ with an estimate of $g(0)/g(z)$) in blue solid line along with the fitted AAR adjusted for thinning (obtained as an estimate of $(1+\gamma x)g(0)$) in the purple dotted line, labeled along the top horizontal axis. The age of ice core is labeled along the left vertical axis. The three panels correspond to Dome Fuji, Lake Vostok, and EPICA Dome C (from left to right). The correlation between the AAR and fitted AAR for each segment is also shown in Figure~\ref{Fig3}. 

It is remarkable that adjustment for thinning, which is modeled merely as a decreasing function of age, brings the observed AAR very close to the fitted AAR, which is a linear function of the temperature. This plot makes use of the factor $g(0)$ that represents the nominal rate of accumulation before thinning, when the temperature deviation is zero. Even though this factor is vulnerable to poor estimation \cite{Radhe2022}, any error in that estimation would merely shift the observed and the fitted curves laterally by the same amount, and not alter their relative positions.

Table~\ref{para_est_table} gives the estimated values of $\gamma$ along with the large sample standard error (given in parentheses below the estimate) for the different parts of the three data sets. The parameter $\gamma$ represents the fractional increase in AAR (over and above the rate at long-term average temperature) associated with every $1^\circ$C rise in temperature. 

\begin{table}[h]
    \begin{center}
    \caption{Estimate of the parameter $\gamma$ of model~\eqref{eqn1} along with error estimates}\label{para_est_table}
    \medskip
        \begin{tabular}{llccc}
        \hline
        Site & Attribute&\multicolumn{3}{@{\hskip3pt}c}{Effect of temperature (per ${}^\circ$C)}\\
        \cline{3-5}
        &&Ages prior to&Ages between&Ages after\\
        &&125 KYrs&125 to 410 KYrs&410 KYrs\\
        \hline
        Dome&Estimated $\gamma$&0.0496 &0.0784 &0.0640 \\
        Fuji&Estimated se & 0.0090&0.0045 &0.0117\\
        &95\% CI& (0.0384, 0.0741)&
        (0.0680, 0.0862)&
        (0.0405, 0.0855)\\[1ex]
        Lake&Estimated $\gamma$&0.0586 &0.0562 &\\
        Vostok&Estimated se& 0.0092&0.0021&\\
        &95\% CI&(0.0413, 0.0762)&(0.0502, 0.0583)&\\[1ex]
        EPICA&Estimated $\gamma$&0.0556 &0.0597 &0.0726 \\
        Dome C&Estimated se&0.0027&0.0015&0.0054\\
        &95\% CI&(0.0508, 0.0613)&(0.0558,0.0619)&(0.0606, 0.0813)\\
    \hline
    \end{tabular}\\
    \end{center}
\end{table}

The positive sign of estimated $\gamma$ in all the analyses indicates local fluctuations in the apparent accumulation rate moving in the direction of fluctuations in temperature. The estimated $\gamma$ for the different sections of the three data sets lie in between 0.050 and 0.078. The variation in the estimated $\gamma$ across different segments may be due to various factors other than temperature. The presence of green house gases and aerosols other than water molecules may affect AAR. 

A closer look at the correlation between AAR and fitted AAR, shown in Figure~\ref{Fig3} reveals that there is quite a lot of variation in the degree of fit in the different segments of the data. The correlations are larger in the middle segment, which is commensurate with the narrower confidence limits of $\gamma$ in that segment. The weaker fit in the other segments indicates the presence of factors other than temperature fluctuations. Effect of the substrata below the lower ice layer is a factor governing the thickness of old accumulate lying at the lower layers, which may have affected the fit at older ages. The top segment has a faster rate of thinning and is more susceptible to disturbance from lateral flow.

The higher elevation of Dome Fuji (see Table~\ref{DATA_info}) may be a factor in governing the effect of temperature. Variation in elevation is known to contribute significantly to inter-site differences in accumulation \cite{Masson-Delmotte2011}. Another possible factor is the surface topography of Dome Fuji, which may be conducive to promoting the lateral flow of ice away from the central pile. The station is placed above a flat bedrock about 800 m in elevation with a structural feature exhibiting a saddle point at the Dome summit. The overall topography is represented by a hill with elevation of ~1000 m a.s.l. positioned at the southeast and northwestern end, and lowland topography of 600 m a.s.l. featured in the west \cite{Watanabe2003}.

The excellent fit at Lake Vostok and EPICA Dome C between 125 and 410 KYrBP, indicating near absence of unaccounted factors, make the two corresponding segments of data ideal for estimating the exclusive effect of temperature. The estimates of $\gamma$ for these segments are 0.056 and 0.060, which correspond to an increment of 5.6\%\ and 6.0\%\, respectively, in the AAR (over and above the accumulation rate at long-term average temperature) associated with every $1^\circ$C increase in temperature. This finding may be appreciated in the context of the widely accepted estimate of about 7\%\ increase in the atmospheric moisture holding capacity per degree Celsius of rise in ambient temperature, as mentioned in Section 1 \cite{Allan2008,Algarra2020,Allan2022}. Greater amount of water vapour in the atmosphere circulating above the surface inversion layer is indeed expected to lead to greater precipitation \cite{Robin1977}. The remarkable part of the present finding is that the AAR adjusted for thinning represents the net accumulation on the ground after accounting for contemporaneous changes through lateral transport and loss through sublimation, melting, evaporation, etc. This fact strongly suggests resilience of the East Antarctic inland ice cap in a warming climate, and corroborates the findings of \cite{Shakun2018}.

Further analysis, the details of which are not reported here, indicated a nonlinear effect of temperature. Specifically, observed accumulation at both ends of observed the range of temperature were marginally higher than the linear fit.

As a caveat, it may be added that `temperature' in this analysis should be interpreted as its proxy obtained from the change in oxygen isotope concentration. There is need for reconciliation of the findings with altimetric studies of recent years and study of implications to the net annual gain or loss of ice in Antarctica.

\subsection{Thinning pattern of AAR}\label{SS2.2}

Figure~\ref{Fig2} reveals a consistent pattern of gradual thinning of ice. Faster thinning in the recent past, due to a greater scope of compaction, is generally followed by a period of slower thinning. 
Interestingly the three plots of AAR in log scale exhibit a linearly decreasing pattern for the age of ice in the middle segment of data (age range 125 to 410 KYrBP) for all the three sites chosen over the region of Antarctica. This linear decline in the log scale is indicative of an exponentially decaying pattern of thinning. Further, for Dome Fuji and EPICA Dome C, where longer records are available, the general level of AAR eventually saturates to a minimum value. This trend is captured by the following parametric version of the model~\eqref{eqn1}: 
\begin{equation}
b(x,z)\approx (1+\gamma x)(b_0 e^{-z/ z_e} + b_\infty) \qquad \mbox{for }125<z<800.
\label{eqn2}
\end{equation}
In this model, 
the parameter $z_e$ is a time constant that controls the rate of reduction of AAR, $b_0+b_\infty$ and $b_\infty$ are the initial and the saturated values of AAR when the temperature deviation is zero, and $\gamma$ is as in~\eqref{eqn1}. 

\begin{figure}[h]
\centering
\begin{tabular}{ccc}
{\underline{\small Dome Fuji}}& {\underline{\small Lake Vostok}}& {\underline{\small EPICA Dome C}}\\
\includegraphics[width=1.4in,height=1.9in]{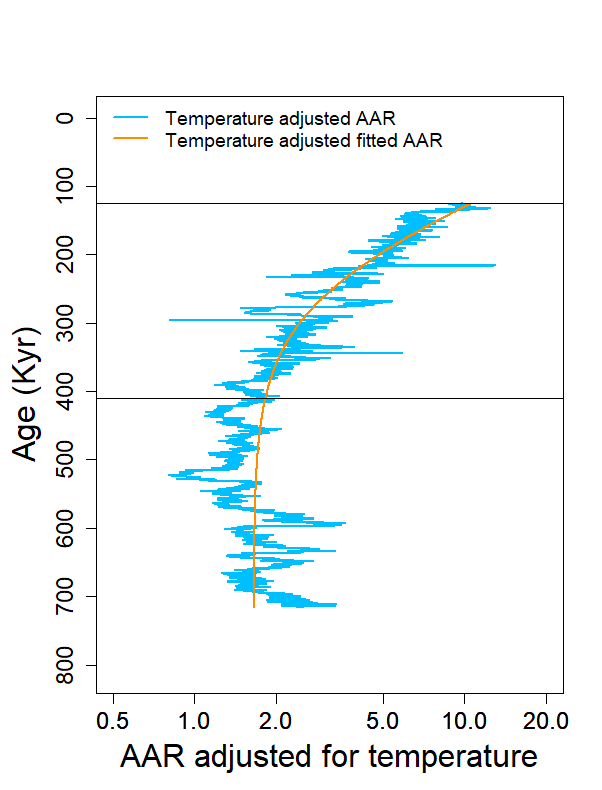}&
\includegraphics[width=1.4in,height=1.9in]{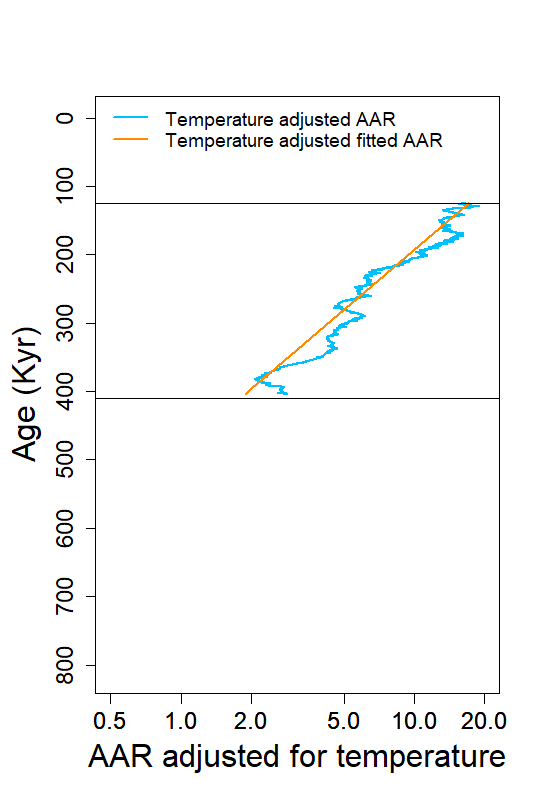}&
\includegraphics[width=1.4in,height=1.9in]{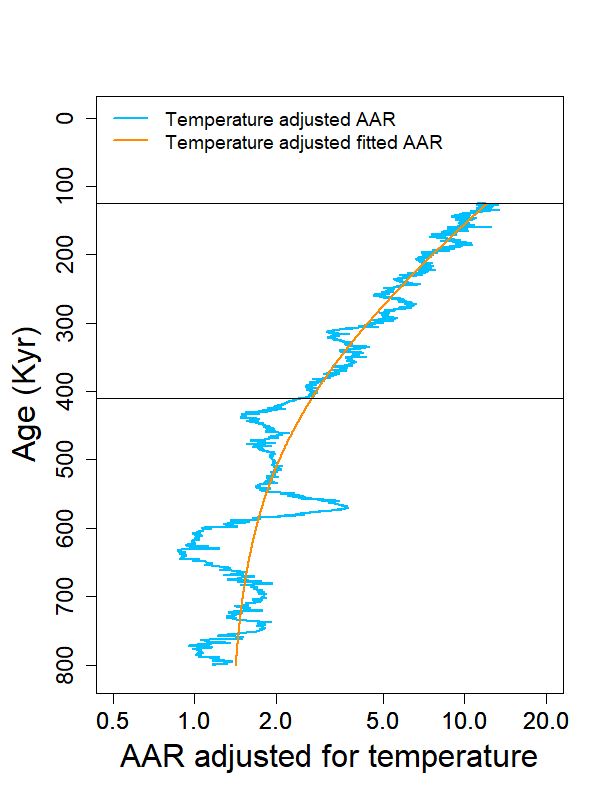}\\
\end{tabular}
\caption{The plots show the graphs of observed AAR adjusted for temperature (in Solid blue line) and fitted AAR adjusted for temperature (in solid orange line) in log scale labeled along the bottom horizontal axis versus age labeled along the left vertical axis for the ice age in the range 125 KYrBP to 800 KYrBP}
\label{Fig5}
\end{figure}

Figure~\ref{Fig5} shows the plots of observed AAR, adjusted for variation in temperature (obtained by dividing AAR with estimated value of $(1+\gamma x)$), along with fitted value AAR, also adjusted for variation in temperature (obtained as the estimate of $(b_0 e^{-z/ z_e} + b_\infty)$), versus age $z$ in solid blue and orange lines, respectively. The adjusted AAR is labeled in the log scale along the bottom horizontal axis and age is labeled on the left vertical axis. The fit is quite good. 

Table~\ref{para_est_table2} shows the estimates of parameters of model~\eqref{eqn2}, along with estimates of the long term thinning factor $b_\infty/(b_0+b_\infty)$. The standard errors of all the estimates, obtained under the assumption of large sample size \cite{White1984}, are shown below each estimate in parentheses. The details of the model and methodology for arriving at these numbers are given in Section~\ref{Sec3}. The parameter $b_\infty$ and $b_\infty/(b_0+b_\infty)$ for Lake Vostok are small with a large standard error and are statistically insignificant.  This is not surprising, as the data record at Lake Vostok covers a much shorter time period. 
\begin{table}[h]
    \begin{center}
    \caption{Parameter estimates of AAR model~\eqref{eqn2} along with estimated standard error}\label{para_est_table2}
    \medskip
        \begin{tabular}{lccccc}
        \hline
        &$\gamma$&$z_e$&$b_0$&$b_\infty$&$\displaystyle\frac{b_\infty}{b_0+b_\infty}$\\
        Location&(S.E.)&(S.E.)&(S.E.)&(S.E.)&(S.E.)\\
        &(per ${}^\circ$C)&(kilo years)&(mm/year)&(mm/year)\\
        \hline
        Dome& 0.0687 &72.470&49.917&1.658&0.0322\\
        Fuji& (0.0065)& (6.684)&(14.260)&(0.055)&(0.0085)\\[1ex]
        Lake&0.0508 & 126.266& 46.203&0.547& 0.0117\\
        Vostok& (0.0052)& (26.694)&(9.639)& (1.180)&(0.0230)\\[1ex]
        EPICA& 0.0626 &138.635&26.540&1.342&0.0481\\
        Dome C&(0.0036)&(14.547)&(4.075)&(0.170)&(0.0059)\\
    \hline
    %\multicolumn{3}{l}{${}^*$value constrained to zero.}&\\
    \end{tabular}
    \end{center}
\end{table}

Saturation of the thinning process is not prominent in the deepest layers of the ice sheet at Lake Vostok, due to the presence of water in the subglacial lake below the drilling site. Observed patterns of differential melting and freezing at the base of the ice sheet \cite{Siegert2000} and flow of ice in the subglacial lake \cite{Leonard2004} indicate significant quantities of water/heat exchange.

Apart from this anomaly at Lake Vostok, the estimated parameters at the three sites follow a similar pattern. The estimates of $\gamma$ are consistent with the range of values obtained from the model~\eqref{eqn1} and reported in Table~\ref{para_est_table} for smaller segments of age.

The initial rate of decline in AAR, plotted in the log scale in Figure~\ref{Fig5}, is very much linear at all the three locations, before saturation is approached at Dome Fuji and EPICA Dome C. The smaller value of the time constant $z_e$ at Dome Fuji is indicative of faster thinning, possibly due to lateral transport away from the site over the ages owing to its higher elevation (see Table~\ref{DATA_info}). The smaller value of the long term thinning factor may also have a similar explanation.

The estimates of the initial AAR $b_0+b_\infty$ at Dome Fuji and Lake Vostok are somewhat higher than the contemporary precipitation records. This deviation is explained by the fact that the parameter is extrapolated from data records older than 125 KYrBP. AAR is the combined effect of precipitation, lateral flow and other prevailing factors during this period.

\section{Models and methods}\label{Sec3}
The AAR over a particular age interval is the ratio between the difference of depths of successive ice-core slices and the age difference of those slices, namely,
\begin{equation*}
b_i = \frac{d_i-d_{i-1}}{a_i-a_{i-1}},   \qquad \mbox{for } i = 2, 3,\cdots  , n,
\end{equation*}
where $d_i$ and $a_i$ are the depth and the age, respectively, of the $i^{\rm th}$ observation of the ice core for  $i = 1,2,...,n$, and $n$ is the total number of slices. We would like to relate the AAR ($b_i$) to the average temperature anomaly during the ages $a_{i-1}$ to $a_i$ over and above its long term average (henceforth referred to as ‘temperature’ and denoted as $x_i$), and the average of the ages $a_{i-1}$ and $a_i$ (henceforth referred to as ‘age’ and denoted as $z_i$). A formal description of the model~\eqref{eqn1} for AAR $b_i$ over the range of depths $d_{i-1}$ to $d_i$ is
\begin{equation}
b_i= (1+\gamma x_i)g(z_i)\xi_i, \qquad \mbox{for } i = 2, 3, \ldots, n.
\label{eqnlinear}
\end{equation}
In this semi-parametric model, $\gamma$ is a constant parameter and $g$ is a decreasing function, and the factor $\xi_i$ is a random and multiplicative error. 

In contrast with the theoretical models available for ice accumulation rate \cite{Leysinger2004,Macgregor2009,Li2011}, the model~\eqref{eqnlinear} proposed here is an empirical one (Figure~\ref{Fig2}) and is supported by physical interpretation. The first factor on the right-hand side of~\eqref{eqnlinear} is a linear function of the average temperature, with slope~$\gamma$. The second factor is what the AAR is expected to be at age $z_i$, if the temperature stayed constant at the long term average. This function is meant to capture the long-term decreasing trend of the AAR as seen in Figure~\ref{Fig2}, which denote an aggregate thinning of ice over time (through relocation/expulsion of gas and water molecules trapped in ice and compaction under the weight of the bulk of ice lying above). The AICC2012 age scale is chosen with an eye on the delineation between the two factors of \eqref{eqnlinear}, as this scale makes minimal use of the information contained in the temperature data.  The first factor is sensitive to variation in temperature, which may affect not only the rate of precipitation but also lateral transport and loss through sublimation, melting or evaporation. The second factor embodies the densification of porous ice and other changes that take place long after its deposition. A fundamental knowledge of that process is important for establishing an accurate age of a strata of ice, and in particular for estimating the time lag between the (older) age of the firn at the time of bubble closure at depth and the (younger) age of CO${}_2$ and other gas trapped within the bubbles.

If AAR is known at frequent intervals of depth, then the age is obtained by integrating the reciprocal of AAR with respect to the depth. Modeling of AAR as the product of two factors as in~\eqref{eqnlinear} implies that the age is the integral of the reciprocal of that product over the relevant range of depths. A similar model was used in \cite{Parrenin2004} for age determination in an ice core study, where the second factor was referred to as the thinning factor. 

The model~\eqref{eqnlinear} is fitted to the data after a log transformation of AAR, which makes the random disturbance additive, and minimization of a least squares criterion through an iterative procedure that alternates between optimization over the parametric part (effect of temperature) and the nonparametric part (corresponding to thinning). The standard errors reported in Table~\ref{para_est_table} are obtained from a model based bootstrap procedure~(see \cite{Radhe2022} for details of assumptions, results, and computation).

Two points of strength of the model~\eqref{eqnlinear} need highlighting. Since the decreasing function $g$ is completely unspecified, it can absorb any error in the estimation of the time scale as long as the estimated age is an increasing function of the actual age. Further, estimation of $\gamma$ is interlinked with the estimation of the entire function $g$, and not with that of $g(0)$ alone. This fact allows our inference on $\gamma$ to be resilient to the well known problems of a nonparametric estimator of a function near the boundaries~\cite{Hardle1990}. In other words, our inference on the effect of temperature benefits from the lack of distinction between the accumulation rate and the thinning process.

A formal version of the fully parametric model~\eqref{eqn2} for AAR $b_i$ over the range of depths $d_{i-1}$ to $d_i$, corresponding to average age $z_i$ and temperature deviation $x_i$, is 
\begin{equation}
b_i= (1+\gamma x_i)(b_0 e^{-z_i/z_e} + b_\infty)\xi_i \qquad \mbox{for } 125<z_i<800,
\label{expdecay}
\end{equation}
where $b_\infty$ is the saturated value of AAR at long term average temperature, $b_0$ is the initial excess value of AAR over $b_\infty$ at long term average temperature, $z_e$ is the time in kiloyears needed by the excess of AAR over $b_\infty$ to reduce by a factor of $e$, and $\xi_i$ is a random and multiplicative disturbance factor representing the effects other causes. The parameter $\gamma$ is the fractional increase in AAR for $1^\circ$C rise in temperature. The ratio $b_\infty/(b_0+b_\infty)$ reported in Table~\ref{para_est_table2} represents the ratio of the long-run AAR with the AAR at the present time, as explained by the parametric model. We refer to this ratio as the long-term thinning factor. The fitting of this model is done by using the method of nonlinear least squares for the log-transformed AAR. The standard errors reported in Table~\ref{para_est_table2} are computed 
from the expressions given in the Appendix, obtained 
by using large sample properties of this estimator given in Theorem 3.2 of~\cite{White1984}.

\section{Data and code availability}\label{Sec4}

The data sets used for the analysis is downloaded from the website of National Center for Environmental Information. The ice core data set for Dome Fuji is available at 
\url{https://www.ncei.noaa.gov/pub/data/paleo/icecore/antarctica/domefuji/domefuji2018iso-temp.txt}. The ice core data sets for Lake Vostok and EPICA Dome C are downloaded from \url{https://www.ncei.noaa.gov/pub/data/paleo/icecore/antarctica/vostok/deutnat.txt} and \url{https://www.ncei.noaa.gov/pub/data/paleo/icecore/antarctica/epica_domec/edc3deuttemp2007.txt
}, respectively. The age scale of ice cores for Lake Vostok and EPICA Dome C is GT4 and EDC3, respectively, as described in the website. 
The AICC2012 age scales~\cite{Bazin2013} for Lake Vostok and EPICA Dome C are downloaded from
\url{https://www.ncei.noaa.gov/access/paleo-search/study/15076} (aicc2012icecore-data.xls). The AICC2012 age scale is used for all the locations for comparability.

The data analysis is performed by using the R software. The code is available on request.

\section*{Acknowledgment} 
We thank Prof. Leuenberger for his valuable comments that led to several improvements in the article.  

\section*{Appendix: Standard errors in Table~\ref{para_est_table2}}\label{Sec5}
The nonlinear least squares estimator $(\hat{\gamma},\hat{b}_0,\hat{z}_e,\hat{b}_\infty)'$ minimizes the sum of squared errors
\begin{eqnarray*}
\sum_{i: 125<z_i<800} \left(\log(b_i)-\log(1+\gamma x_i)-\log(b_0 e^{-z_i/z_e} + b_\infty)\right)^2.
\end{eqnarray*}
Under the conditions of Theorem 3.2 of~\cite{White1984} and the additional assumption that $\log(\xi_i)$ is an autoregressive process of order 1 (as in \cite{Radhe2022}), a consistent estimator of the covariance matrix of the estimator $(\hat{\gamma},\hat{b}_0,\hat{z}_e,\hat{b}_\infty)'$ is given by
\begin{equation*}
\hat{V}=\frac1n (\hat{F}'\hat{F}/n)^{-1}(\hat{F}'\hat{\Sigma} \hat{F}/n) (\hat{F}'\hat{F}/n)^{-1},
\end{equation*}
where $n$ is the sample size, $\hat{F}$ is a $n\times 4$ matrix with $i^{th}$ row 
\begin{eqnarray*}
    \begin{bmatrix}
     \frac{x_i}{1+\hat{\gamma} x_i}&
     \frac{e^{-z_i/\hat{z}_e}}{\hat{b}_0e^{-z_i/\hat{z}_e}+\hat{b}_\infty}&
     \frac{-\hat{b}_0 z_i e^{-z_i/\hat{z}_e}}{\hat{z}_e^2(\hat{b}_0e^{-z_i/\hat{z}_e}+\hat{b}_\infty)}&
     \frac{1}{\hat{b}_0e^{-z_i/\hat{z}_e}+\hat{b}_\infty}
    \end{bmatrix},
\end{eqnarray*}
and $\hat{\Sigma}$ is the $n\times n$ Toeplitz matrix with $(i,j)^{\rm th}$ element %generated from the vector 
\begin{equation*}
    \hat s_{ij}=\frac{\hat{\sigma}^2}{1-\hat{\phi}^2}
    \hat{\phi}^{\mid i-j\mid}
%    \begin{bmatrix} 1& \hat{\phi} & \hat{\phi}^2&\cdots&\hat{\phi}^{n-1} \end{bmatrix}
%\end{eqnarray*}
\qquad\mbox{with } 
%\begin{eqnarray*}
\hat{\phi}=\frac{\frac1n\sum_{i=2}^n\hat{\varepsilon}_i\hat{\varepsilon}_{i-1}}{\frac1n\sum_{i=2}^n\hat{\varepsilon}_i^2},\ %&\mbox{and}&
\hat{\sigma}^2=\frac1n \sum_{i=2}^n \left(\hat{\varepsilon}_i -\hat{\phi} \hat{\varepsilon}_{i-1}\right)^2,
\end{equation*}
and
\begin{eqnarray*}
\hat{\varepsilon}_i&=&\log(b_i)-\log (1+\hat{\gamma} x_i)-\log(\hat{b}_0 e^{-z_i/\hat{z}_e} + \hat{b}_\infty), \mbox{ for } i=1,2,\ldots,n.
\end{eqnarray*}
The standard error of $b_{\infty}/(b_\infty+b_0)$, obtained by using the Delta method, is given by
$$\frac{1}{(\hat{b}_\infty+\hat{b}_0)^2}\sqrt{\hat{b}_0^2 \ v_{4,4}+
\hat{b}_\infty^2 \  v_{2,2}-
2\hat{b}_0\hat{b}_\infty \  v_{2,4}} \ \ ,$$
where $v_{i,j}$ is the $(i,j)^{\mbox{th}}$ element of matrix $\hat{V}$ for $i,j=1,2,3,4$.

\bibliography{sn-bibliography}

%% BioMed_Central_Bib_Style_v1.01

\begin{thebibliography}{32}
% BibTex style file: bmc-mathphys.bst (version 2.1), 2014-07-24
\ifx \bisbn   \undefined \def \bisbn  #1{ISBN #1}\fi
\ifx \binits  \undefined \def \binits#1{#1}\fi
\ifx \bauthor  \undefined \def \bauthor#1{#1}\fi
\ifx \batitle  \undefined \def \batitle#1{#1}\fi
\ifx \bjtitle  \undefined \def \bjtitle#1{#1}\fi
\ifx \bvolume  \undefined \def \bvolume#1{\textbf{#1}}\fi
\ifx \byear  \undefined \def \byear#1{#1}\fi
\ifx \bissue  \undefined \def \bissue#1{#1}\fi
\ifx \bfpage  \undefined \def \bfpage#1{#1}\fi
\ifx \blpage  \undefined \def \blpage #1{#1}\fi
\ifx \burl  \undefined \def \burl#1{\textsf{#1}}\fi
\ifx \doiurl  \undefined \def \doiurl#1{\url{https://doi.org/#1}}\fi
\ifx \betal  \undefined \def \betal{\textit{et al.}}\fi
\ifx \binstitute  \undefined \def \binstitute#1{#1}\fi
\ifx \binstitutionaled  \undefined \def \binstitutionaled#1{#1}\fi
\ifx \bctitle  \undefined \def \bctitle#1{#1}\fi
\ifx \beditor  \undefined \def \beditor#1{#1}\fi
\ifx \bpublisher  \undefined \def \bpublisher#1{#1}\fi
\ifx \bbtitle  \undefined \def \bbtitle#1{#1}\fi
\ifx \bedition  \undefined \def \bedition#1{#1}\fi
\ifx \bseriesno  \undefined \def \bseriesno#1{#1}\fi
\ifx \blocation  \undefined \def \blocation#1{#1}\fi
\ifx \bsertitle  \undefined \def \bsertitle#1{#1}\fi
\ifx \bsnm \undefined \def \bsnm#1{#1}\fi
\ifx \bsuffix \undefined \def \bsuffix#1{#1}\fi
\ifx \bparticle \undefined \def \bparticle#1{#1}\fi
\ifx \barticle \undefined \def \barticle#1{#1}\fi
\bibcommenthead
\ifx \bconfdate \undefined \def \bconfdate #1{#1}\fi
\ifx \botherref \undefined \def \botherref #1{#1}\fi
\ifx \url \undefined \def \url#1{\textsf{#1}}\fi
\ifx \bchapter \undefined \def \bchapter#1{#1}\fi
\ifx \bbook \undefined \def \bbook#1{#1}\fi
\ifx \bcomment \undefined \def \bcomment#1{#1}\fi
\ifx \oauthor \undefined \def \oauthor#1{#1}\fi
\ifx \citeauthoryear \undefined \def \citeauthoryear#1{#1}\fi
\ifx \endbibitem  \undefined \def \endbibitem {}\fi
\ifx \bconflocation  \undefined \def \bconflocation#1{#1}\fi
\ifx \arxivurl  \undefined \def \arxivurl#1{\textsf{#1}}\fi
\csname PreBibitemsHook\endcsname

%%% 1
\bibitem{Rapp2019}
\begin{bbook}
\bauthor{\bsnm{Rapp}, \binits{D.}}:
\bbtitle{Ice Core Data},
pp. \bfpage{83}--\blpage{118}.
\bpublisher{Springer},
\blocation{Cham}
(\byear{2019})
\end{bbook}
\endbibitem

%%% 2
\bibitem{Jouzel2007}
\begin{botherref}
\oauthor{\bsnm{{Jouzel}}, \binits{J.}},
\oauthor{\bsnm{{Masson-Delmotte}}, \binits{V.}}:
{EPICA Dome C Ice Core 800KYr deuterium data and temperature estimates}.
PANGAEA.
Supplement to: Jouzel, Jean et al. (2007): Orbital and millennial Antarctic climate variability over the past 800,000 years. Science, 317(5839), 793-797
(2007)
\end{botherref}
\endbibitem

%%% 3
\bibitem{Petit1999}
\begin{barticle}
\bauthor{\bsnm{Petit}, \binits{J.R.}},
\bauthor{\bsnm{Jouzel}, \binits{J.}},
\bauthor{\bsnm{Raynaud}, \binits{D.}},
\bauthor{\bsnm{Barkov}, \binits{N.I.}},
\bauthor{\bsnm{Barnola}, \binits{J.-M.}},
\bauthor{\bsnm{Basile}, \binits{I.}},
\bauthor{\bsnm{Bender}, \binits{M.}},
\bauthor{\bsnm{Chappellaz}, \binits{J.}},
\bauthor{\bsnm{Davis}, \binits{M.}},
\bauthor{\bsnm{Delaygue}, \binits{G.}},
\bauthor{\bsnm{Delmotte}, \binits{M.}},
\bauthor{\bsnm{Kotlyakov}, \binits{V.M.}},
\bauthor{\bsnm{Legrand}, \binits{M.}},
\bauthor{\bsnm{Lipenkov}, \binits{V.Y.}},
\bauthor{\bsnm{Lorius}, \binits{C.}},
\bauthor{\bsnm{P\'{E}pin}, \binits{L.}},
\bauthor{\bsnm{Ritz}, \binits{C.}},
\bauthor{\bsnm{Saltzman}, \binits{E.}},
\bauthor{\bsnm{Stievenard}, \binits{M.}}:
\batitle{Climate and atmospheric history of the past 420,000 years from the {V}ostok ice core, {A}ntarctica}.
\bjtitle{Nature}
\bvolume{399}(\bissue{6735}),
\bfpage{429}--\blpage{436}
(\byear{1999})
\end{barticle}
\endbibitem

%%% 4
\bibitem{Siegenthaler1987}
\begin{barticle}
\bauthor{\bsnm{Siegenthaler}, \binits{U.}},
\bauthor{\bsnm{Oeschger}, \binits{H.}}:
\batitle{Biospheric {CO2} emissions during the past 200 years reconstructed by deconvolution of ice core data}.
\bjtitle{Tellus B: Chemical and Physical Meteorology}
\bvolume{39}(\bissue{1-2}),
\bfpage{140}--\blpage{154}
(\byear{1987})
\end{barticle}
\endbibitem

%%% 5
\bibitem{Algarra2020}
\begin{barticle}
\bauthor{\bsnm{Algarra}, \binits{I.}},
\bauthor{\bsnm{Nieto}, \binits{R.}},
\bauthor{\bsnm{Ramos}, \binits{A.M.}},
\bauthor{\bsnm{Eiras-Barca}, \binits{J.}},
\bauthor{\bsnm{Trigo}, \binits{R.M.}},
\bauthor{\bsnm{Gimeno}, \binits{L.}}:
\batitle{Significant increase of global anomalous moisture uptake feeding landfalling atmospheric rivers}.
\bjtitle{Nature {C}ommunications}
\bvolume{11}(\bissue{1}),
\bfpage{2041}--\blpage{1723}
(\byear{2020})
\end{barticle}
\endbibitem

%%% 6
\bibitem{Thomas2008}
\begin{botherref}
\oauthor{\bsnm{Thomas}, \binits{E.R.}},
\oauthor{\bsnm{Marshall}, \binits{G.J.}},
\oauthor{\bsnm{McConnel}, \binits{J.R.}}:
A doubling in snow accumulation in the western antarctic peninsula since 1850.
Geophysical Research Letters
\textbf{35}(1)
(2008)
\end{botherref}
\endbibitem

%%% 7
\bibitem{vandenBroeke2006}
\begin{botherref}
\oauthor{\bparticle{van~den} \bsnm{Broeke}, \binits{M.}},
\oauthor{\bparticle{van~de} \bsnm{Berg}, \binits{W.J.}},
\oauthor{\bparticle{van} \bsnm{Meijgaard}, \binits{E.}}:
Snowfall in coastal {W}est {A}ntarctica much greater than previously assumed.
Geophysical Research Letters
\textbf{33}(2)
(2006)
\end{botherref}
\endbibitem

%%% 8
\bibitem{Wingham2006}
\begin{barticle}
\bauthor{\bsnm{Wingham}, \binits{D.J.}},
\bauthor{\bsnm{Shepherd}, \binits{A.}},
\bauthor{\bsnm{Muir}, \binits{A.}},
\bauthor{\bsnm{Marshall}, \binits{G.J.}}:
\batitle{Mass balance of the {A}ntarctic ice sheet}.
\bjtitle{Philosophical Transactions of the Royal Society A: Mathematical, Physical and Engineering Sciences}
\bvolume{364}(\bissue{1844}),
\bfpage{1627}--\blpage{1635}
(\byear{2006})
\end{barticle}
\endbibitem

%%% 9
\bibitem{Yang2006}
\begin{barticle}
\bauthor{\bsnm{Yang}, \binits{M.}},
\bauthor{\bsnm{Yao}, \binits{T.}},
\bauthor{\bsnm{Wang}, \binits{H.}},
\bauthor{\bsnm{Gou}, \binits{X.}}:
\batitle{Correlation between precipitation and temperature variations in the past 300 years recorded in {G}uliya ice core, {C}hina}.
\bjtitle{Annals of Glaciology}
\bvolume{43},
\bfpage{137}--\blpage{141}
(\byear{2006})
\end{barticle}
\endbibitem

%%% 10
\bibitem{Fudge2016}
\begin{barticle}
\bauthor{\bsnm{Fudge}, \binits{T.J.}},
\bauthor{\bsnm{Markle}, \binits{B.R.}},
\bauthor{\bsnm{Cuffey}, \binits{K.M.}},
\bauthor{\bsnm{Buizert}, \binits{C.}},
\bauthor{\bsnm{Taylor}, \binits{K.C.}},
\bauthor{\bsnm{Steig}, \binits{E.J.}},
\bauthor{\bsnm{Waddington}, \binits{E.D.}},
\bauthor{\bsnm{Conway}, \binits{H.}},
\bauthor{\bsnm{Koutnik}, \binits{M.}}:
\batitle{Variable relationship between accumulation and temperature in {W}est {A}ntarctica for the past 31,000 years}.
\bjtitle{Geophysical Research Letters}
\bvolume{43}(\bissue{8}),
\bfpage{3795}--\blpage{3803}
(\byear{2016})
\end{barticle}
\endbibitem

%%% 11
\bibitem{Cuffey1997}
\begin{barticle}
\bauthor{\bsnm{Cuffey}, \binits{K.M.}},
\bauthor{\bsnm{Clow}, \binits{G.D.}}:
\batitle{Temperature, accumulation, and ice sheet elevation in central {G}reenland through the last deglacial transition}.
\bjtitle{Journal of Geophysical Research: Oceans}
\bvolume{102}(\bissue{C12}),
\bfpage{26383}--\blpage{26396}
(\byear{1997})
\end{barticle}
\endbibitem

%%% 12
\bibitem{Cook2013}
\begin{barticle}
\bauthor{\bsnm{Cook}, \binits{C.P.}},
\bauthor{\bparticle{van~de} \bsnm{Flierdt}, \binits{T.}},
\bauthor{\bsnm{Williams}, \binits{T.}},
\bauthor{\bsnm{Hemming}, \binits{S.R.}},
\bauthor{\bsnm{Iwai}, \binits{M.}},
\bauthor{\bsnm{Kobayashi}, \binits{M.}},
\bauthor{\bsnm{Jimenez-Espejo}, \binits{F.J.}},
\bauthor{\bsnm{Escutia}, \binits{C.}},
\bauthor{\bsnm{Gonz\'{a}lez}, \binits{J.J.}},
\bauthor{\bsnm{Khim}, \binits{B.-K.}},
\bauthor{\bsnm{McKay}, \binits{R.M.}},
\bauthor{\bsnm{Passchier}, \binits{S.}},
\bauthor{\bsnm{Bohaty}, \binits{S.M.}},
\bauthor{\bsnm{Riesselman}, \binits{C.R.}},
\bauthor{\bsnm{Tauxe}, \binits{L.}},
\bauthor{\bsnm{Sugisaki}, \binits{S.}},
\bauthor{\bsnm{Galindo}, \binits{A.L.}},
\bauthor{\bsnm{Patterson}, \binits{M.O.}},
\bauthor{\bsnm{Sangiorgi}, \binits{F.}},
\bauthor{\bsnm{Pierce}, \binits{E.L.}},
\bauthor{\bsnm{Brinkhuis}, \binits{H.}},
\bauthor{\bsnm{Klaus}, \binits{A.}},
\bauthor{\bsnm{Fehr}, \binits{A.}},
\bauthor{\bsnm{Bendle}, \binits{J.A.P.}},
\bauthor{\bsnm{Bijl}, \binits{P.K.}},
\bauthor{\bsnm{Carr}, \binits{S.A.}},
\bauthor{\bsnm{Dunbar}, \binits{R.B.}},
\bauthor{\bsnm{Flores}, \binits{J.A.}},
\bauthor{\bsnm{Hayden}, \binits{T.G.}},
\bauthor{\bsnm{Katsuki}, \binits{K.}},
\bauthor{\bsnm{Kong}, \binits{G.S.}},
\bauthor{\bsnm{Nakai}, \binits{M.}},
\bauthor{\bsnm{Olney}, \binits{M.P.}},
\bauthor{\bsnm{Pekar}, \binits{S.F.}},
\bauthor{\bsnm{Pross}, \binits{J.}},
\bauthor{\bsnm{R{\"o}hl}, \binits{U.}},
\bauthor{\bsnm{Sakai}, \binits{T.}},
\bauthor{\bsnm{Shrivastava}, \binits{P.K.}},
\bauthor{\bsnm{Stickley}, \binits{C.E.}},
\bauthor{\bsnm{Tuo}, \binits{S.}},
\bauthor{\bsnm{Welsh}, \binits{K.}},
\bauthor{\bsnm{Yamane}, \binits{M.}}:
\batitle{Dynamic behaviour of the {E}ast {A}ntarctic ice sheet during pliocene warmth}.
\bjtitle{Nature {G}eoscience}
\bvolume{6}(\bissue{9}),
\bfpage{765}--\blpage{769}
(\byear{2013})
\end{barticle}
\endbibitem

%%% 13
\bibitem{Robinson2008}
\begin{barticle}
\bauthor{\bsnm{Robinson}, \binits{M.M.}},
\bauthor{\bsnm{Dowsett}, \binits{H.J.}},
\bauthor{\bsnm{Chandler}, \binits{M.A.}}:
\batitle{Pliocene role in assessing future climate impacts}.
\bjtitle{Eos, Transactions American Geophysical Union}
\bvolume{89}(\bissue{49}),
\bfpage{501}--\blpage{502}
(\byear{2008})
\end{barticle}
\endbibitem

%%% 14
\bibitem{delaVega2020}
\begin{barticle}
\bauthor{\bparticle{de~la} \bsnm{Vega}, \binits{E.}},
\bauthor{\bsnm{Chalk}, \binits{T.B.}},
\bauthor{\bsnm{Wilson}, \binits{P.A.}},
\bauthor{\bsnm{Bysani}, \binits{R.P.}},
\bauthor{\bsnm{Foster}, \binits{G.L.}}:
\batitle{Atmospheric {CO2} during the mid-piacenzian warm period and the {M2} glaciation}.
\bjtitle{Scientific Reports}
\bvolume{10}(\bissue{1}),
\bfpage{11002}
(\byear{2020})
\end{barticle}
\endbibitem

%%% 15
\bibitem{Shakun2018}
\begin{barticle}
\bauthor{\bsnm{Shakun}, \binits{J.D.}},
\bauthor{\bsnm{Corbett}, \binits{L.B.}},
\bauthor{\bsnm{Bierman}, \binits{P.R.}},
\bauthor{\bsnm{Underwood}, \binits{K.}},
\bauthor{\bsnm{Rizzo}, \binits{D.M.}},
\bauthor{\bsnm{Zimmerman}, \binits{S.R.}},
\bauthor{\bsnm{Caffee}, \binits{M.W.}},
\bauthor{\bsnm{Naish}, \binits{T.}},
\bauthor{\bsnm{Golledge}, \binits{N.R.}},
\bauthor{\bsnm{Hay}, \binits{C.C.}}:
\batitle{Minimal {E}ast {A}ntarctic {I}ce {S}heet retreat onto land during the past eight million years}.
\bjtitle{Nature}
\bvolume{558}(\bissue{7709}),
\bfpage{284}--\blpage{287}
(\byear{2018})
\end{barticle}
\endbibitem

%%% 16
\bibitem{Bazin2013}
\begin{botherref}
\oauthor{\bsnm{{Bazin}}, \binits{L.}},
\oauthor{\bsnm{{Landais}}, \binits{A.}},
\oauthor{\bsnm{{Lemieux-Dudon}}, \binits{B.}},
\oauthor{\bsnm{{Toy\'{e} Mahamadou Kele}}, \binits{H.}},
\oauthor{\bsnm{{Veres}}, \binits{D.}},
\oauthor{\bsnm{{Parrenin}}, \binits{F.}},
\oauthor{\bsnm{{Martinerie}}, \binits{P.}},
\oauthor{\bsnm{{Ritz}}, \binits{C.}},
\oauthor{\bsnm{{Capron}}, \binits{E.}},
\oauthor{\bsnm{{Lipenkov}}, \binits{V.Y.}},
\oauthor{\bsnm{{Loutre}}, \binits{M.-F.}},
\oauthor{\bsnm{{Raynaud}}, \binits{D.}},
\oauthor{\bsnm{{Vinther}}, \binits{B.M.}},
\oauthor{\bsnm{{Svensson}}, \binits{A.M.}},
\oauthor{\bsnm{{Rasmussen}}, \binits{S.O.}},
\oauthor{\bsnm{{Severi}}, \binits{M.}},
\oauthor{\bsnm{{Blunier}}, \binits{T.}},
\oauthor{\bsnm{{Leuenberger}}, \binits{M.C.}},
\oauthor{\bsnm{{Fischer}}, \binits{H.}},
\oauthor{\bsnm{{Masson-Delmotte}}, \binits{V.}},
\oauthor{\bsnm{{Chappellaz}}, \binits{J.A.}},
\oauthor{\bsnm{{Wolff}}, \binits{E.W.}}:
{The Antarctic ice core chronology (AICC2012)}.
PANGAEA
(2013)
\end{botherref}
\endbibitem

%%% 17
\bibitem{Cauquoin2015}
\begin{barticle}
\bauthor{\bsnm{Cauquoin}, \binits{A.}},
\bauthor{\bsnm{Landais}, \binits{A.}},
\bauthor{\bsnm{Raisbeck}, \binits{G.M.}},
\bauthor{\bsnm{Jouzel}, \binits{J.}},
\bauthor{\bsnm{Bazin}, \binits{L.}},
\bauthor{\bsnm{Kageyama}, \binits{M.}},
\bauthor{\bsnm{Peterschmitt}, \binits{J.-Y.}},
\bauthor{\bsnm{Werner}, \binits{M.}},
\bauthor{\bsnm{Bard}, \binits{E.}},
\bauthor{\bsnm{Team}, \binits{A.}}:
\batitle{Comparing past accumulation rate reconstructions in east antarctic ice cores using $^{10}$be, water isotopes and cmip5-pmip3 models}.
\bjtitle{Climate of the Past}
\bvolume{11}(\bissue{3}),
\bfpage{355}--\blpage{367}
(\byear{2015})
\end{barticle}
\endbibitem

%%% 18
\bibitem{Kahle2021}
\begin{barticle}
\bauthor{\bsnm{Kahle}, \binits{E.C.}},
\bauthor{\bsnm{Steig}, \binits{E.J.}},
\bauthor{\bsnm{Jones}, \binits{T.R.}},
\bauthor{\bsnm{Fudge}, \binits{T.J.}},
\bauthor{\bsnm{Koutnik}, \binits{M.R.}},
\bauthor{\bsnm{Morris}, \binits{V.A.}},
\bauthor{\bsnm{Vaughn}, \binits{B.H.}},
\bauthor{\bsnm{Schauer}, \binits{A.J.}},
\bauthor{\bsnm{Stevens}, \binits{C.M.}},
\bauthor{\bsnm{Conway}, \binits{H.}},
\bauthor{\bsnm{Waddington}, \binits{E.D.}},
\bauthor{\bsnm{Buizert}, \binits{C.}},
\bauthor{\bsnm{Epifanio}, \binits{J.}},
\bauthor{\bsnm{White}, \binits{J.W.C.}}:
\batitle{Reconstruction of temperature, accumulation rate, and layer thinning from an ice core at south pole, using a statistical inverse method}.
\bjtitle{Journal of Geophysical Research: Atmospheres}
\bvolume{126}(\bissue{13}),
\bfpage{2020}--\blpage{033300}
(\byear{2021})
\end{barticle}
\endbibitem

%%% 19
\bibitem{Radhe2022}
\begin{botherref}
\oauthor{\bsnm{Srivastava}, \binits{R.}},
\oauthor{\bsnm{Sengupta}, \binits{D.}}:
A semi-parametric model for ice accumulation rate and temperature based on {A}ntarctic ice core data.
Preprint at \url{http://arxiv.org/abs/2309.03782}
(2023)
\end{botherref}
\endbibitem

%%% 20
\bibitem{Masson-Delmotte2011}
\begin{barticle}
\bauthor{\bsnm{Masson-Delmotte}, \binits{V.}},
\bauthor{\bsnm{Buiron}, \binits{D.}},
\bauthor{\bsnm{Ekaykin}, \binits{A.}},
\bauthor{\bsnm{Frezzotti}, \binits{M.}},
\bauthor{\bsnm{Gall\'ee}, \binits{H.}},
\bauthor{\bsnm{Jouzel}, \binits{J.}},
\bauthor{\bsnm{Krinner}, \binits{G.}},
\bauthor{\bsnm{Landais}, \binits{A.}},
\bauthor{\bsnm{Motoyama}, \binits{H.}},
\bauthor{\bsnm{Oerter}, \binits{H.}},
\bauthor{\bsnm{Pol}, \binits{K.}},
\bauthor{\bsnm{Pollard}, \binits{D.}},
\bauthor{\bsnm{Ritz}, \binits{C.}},
\bauthor{\bsnm{Schlosser}, \binits{E.}},
\bauthor{\bsnm{Sime}, \binits{L.C.}},
\bauthor{\bsnm{Sodemann}, \binits{H.}},
\bauthor{\bsnm{Stenni}, \binits{B.}},
\bauthor{\bsnm{Uemura}, \binits{R.}},
\bauthor{\bsnm{Vimeux}, \binits{F.}}:
\batitle{A comparison of the present and last interglacial periods in six antarctic ice cores}.
\bjtitle{Climate of the Past}
\bvolume{7}(\bissue{2}),
\bfpage{397}--\blpage{423}
(\byear{2011})
\end{barticle}
\endbibitem

%%% 21
\bibitem{Watanabe2003}
\begin{barticle}
\bauthor{\bsnm{Watanabe}, \binits{O.}},
\bauthor{\bsnm{Kamiyama}, \binits{K.}},
\bauthor{\bsnm{Motoyama}, \binits{H.}},
\bauthor{\bsnm{Fujii}, \binits{Y.}},
\bauthor{\bsnm{Igarashi}, \binits{M.}},
\bauthor{\bsnm{Furukawa}, \binits{T.}},
\bauthor{\bsnm{Goto-Azuma}, \binits{K.}},
\bauthor{\bsnm{Saito}, \binits{T.}},
\bauthor{\bsnm{Kanamori}, \binits{S.}},
\bauthor{\bsnm{Kanamori}, \binits{N.}},
\bauthor{\bsnm{Yoshida}, \binits{N.}},
\bauthor{\bsnm{Uemura}, \binits{R.}}:
\batitle{General tendencies of stable isotopes and major chemical constituents of the {D}ome {F}uji deep ice core}.
\bjtitle{Memoirs of National Institute of Polar Research}
\bvolume{57},
\bfpage{1}--\blpage{24}
(\byear{2003})
\end{barticle}
\endbibitem

%%% 22
\bibitem{Allan2008}
\begin{barticle}
\bauthor{\bsnm{Allan}, \binits{R.P.}},
\bauthor{\bsnm{Soden}, \binits{B.J.}}:
\batitle{Atmospheric warming and the amplification of precipitation extremes}.
\bjtitle{Science}
\bvolume{321}(\bissue{5895}),
\bfpage{1481}--\blpage{1484}
(\byear{2008})
\end{barticle}
\endbibitem

%%% 23
\bibitem{Allan2022}
\begin{barticle}
\bauthor{\bsnm{Allan}, \binits{R.P.}},
\bauthor{\bsnm{Willett}, \binits{K.M.}},
\bauthor{\bsnm{John}, \binits{V.O.}},
\bauthor{\bsnm{Trent}, \binits{T.}}:
\batitle{Global changes in water vapor 1979–2020}.
\bjtitle{Journal of Geophysical Research: Atmospheres}
\bvolume{127}(\bissue{12}),
\bfpage{2022}--\blpage{036728}
(\byear{2022})
\end{barticle}
\endbibitem

%%% 24
\bibitem{Robin1977}
\begin{barticle}
\bauthor{\bsnm{Robin}, \binits{G.D.Q.}},
\bauthor{\bsnm{Mitchell}, \binits{G.F.}},
\bauthor{\bsnm{West}, \binits{R.G.}}:
\batitle{Ice cores and climatic change}.
\bjtitle{Philosophical Transactions of the Royal Society of London. B, Biological Sciences}
\bvolume{280}(\bissue{972}),
\bfpage{143}--\blpage{168}
(\byear{1977})
\end{barticle}
\endbibitem

%%% 25
\bibitem{White1984}
\begin{barticle}
\bauthor{\bsnm{White}, \binits{H.}},
\bauthor{\bsnm{Domowitz}, \binits{I.}}:
\batitle{Nonlinear regression with dependent observations}.
\bjtitle{Econometrica}
\bvolume{52}(\bissue{1}),
\bfpage{143}--\blpage{162}
(\byear{1984})
\end{barticle}
\endbibitem

%%% 26
\bibitem{Siegert2000}
\begin{barticle}
\bauthor{\bsnm{Siegert}, \binits{M.J.}},
\bauthor{\bsnm{Kwok}, \binits{R.}},
\bauthor{\bsnm{Mayer}, \binits{C.}},
\bauthor{\bsnm{Hubbard}, \binits{B.}}:
\batitle{Water exchange between the subglacial {L}ake {V}ostok and the overlying ice sheet}.
\bjtitle{Nature}
\bvolume{403}(\bissue{6770}),
\bfpage{643}--\blpage{646}
(\byear{2000})
\end{barticle}
\endbibitem

%%% 27
\bibitem{Leonard2004}
\begin{barticle}
\bauthor{\bsnm{Leonard}, \binits{K.}},
\bauthor{\bsnm{Bell}, \binits{R.E.}},
\bauthor{\bsnm{Studinger}, \binits{M.}},
\bauthor{\bsnm{Tremblay}, \binits{B.}}:
\batitle{Anomalous accumulation rates in the vostok ice-core resulting from ice flow over {L}ake {V}ostok}.
\bjtitle{Geophysical Research Letters}
\bvolume{31}(\bissue{24}),
\bfpage{24401}
(\byear{2004})
\end{barticle}
\endbibitem

%%% 28
\bibitem{Leysinger2004}
\begin{barticle}
\bauthor{\bsnm{Leysinger~Vieli}, \binits{G.J.-M.C.}},
\bauthor{\bsnm{Siegert}, \binits{M.J.}},
\bauthor{\bsnm{Payne}, \binits{A.J.}}:
\batitle{Reconstructing ice-sheet accumulation rates at ridge {B}, {E}ast {A}ntarctica}.
\bjtitle{Annals of Glaciology}
\bvolume{39},
\bfpage{326}--\blpage{330}
(\byear{2004})
\end{barticle}
\endbibitem

%%% 29
\bibitem{Macgregor2009}
\begin{barticle}
\bauthor{\bsnm{MacGregor}, \binits{J.A.}},
\bauthor{\bsnm{Matsuoka}, \binits{K.}},
\bauthor{\bsnm{Koutnik}, \binits{M.R.}},
\bauthor{\bsnm{Waddington}, \binits{E.D.}},
\bauthor{\bsnm{Studinger}, \binits{M.}},
\bauthor{\bsnm{Winebrenner}, \binits{D.P.}}:
\batitle{Millennially averaged accumulation rates for the {V}ostok subglacial lake region inferred from deep internal layers}.
\bjtitle{Annals of Glaciology}
\bvolume{50}(\bissue{51}),
\bfpage{25}--\blpage{34}
(\byear{2009})
\end{barticle}
\endbibitem

%%% 30
\bibitem{Li2011}
\begin{barticle}
\bauthor{\bsnm{Li}, \binits{J.}},
\bauthor{\bsnm{Zwally}, \binits{H.J.}}:
\batitle{Modeling of firn compaction for estimating ice-sheet mass change from observed ice-sheet elevation change}.
\bjtitle{Annals of Glaciology}
\bvolume{52}(\bissue{59}),
\bfpage{1}--\blpage{7}
(\byear{2011})
\end{barticle}
\endbibitem

%%% 31
\bibitem{Parrenin2004}
\begin{barticle}
\bauthor{\bsnm{Parrenin}, \binits{F.}},
\bauthor{\bsnm{Rémy}, \binits{F.}},
\bauthor{\bsnm{Ritz}, \binits{C.}},
\bauthor{\bsnm{Siegert}, \binits{M.J.}},
\bauthor{\bsnm{Jouzel}, \binits{J.}}:
\batitle{New modeling of the {V}ostok ice flow line and implication for the glaciological chronology of the {V}ostok ice core}.
\bjtitle{Journal of Geophysical Research: Atmospheres}
\bvolume{109}(\bissue{D20}),
\bfpage{20102}
(\byear{2004})
\end{barticle}
\endbibitem

%%% 32
\bibitem{Hardle1990}
\begin{bbook}
\bauthor{\bsnm{H{\"a}rdle}, \binits{W.}}:
\bbtitle{Applied Nonparametric Regression}.
\bpublisher{Cambridge University Press},
\blocation{Cambridge, England}
(\byear{1990})
\end{bbook}
\endbibitem

\end{thebibliography}

\end{document}